\documentclass[aps,pre,twocolumn,superscriptaddress,floatfix]{revtex4-1}

\usepackage{graphicx,bm,amssymb,mathrsfs,amsmath,dcolumn,hyperref}
\usepackage{multirow}
\usepackage{enumerate}
\usepackage{enumitem}
\usepackage{color}
\usepackage{subfigure}  
\usepackage{epstopdf}
\usepackage{bbm}
\usepackage{graphicx}
\usepackage{hyperref}
\usepackage{threeparttable}
\usepackage{amsthm}
\usepackage{mathtools}
\usepackage{tikz}
\usepackage{tikz-3dplot} 
\usepackage{float}
\usepackage{indentfirst}
\usetikzlibrary{math}
\usepackage[utf8]{inputenc} 
\usepackage[greek, main=english]{babel}

\begin{document}
\newcommand{\upcite}[1]{\textsuperscript{\textsuperscript{\cite{#1}}}}
\newcommand{\be}{\begin{equation}}
\newcommand{\ee}{\end{equation}}
\newcommand{\half}{\frac{1}{2}}
\newcommand{\ith}{^{(i)}}
\newcommand{\im}{^{(i-1)}}
\newcommand{\gae}
{\,\hbox{\lower0.5ex\hbox{$\sim$}\llap{\raise0.5ex\hbox{$>$}}}\,}
\newcommand{\lae}
{\,\hbox{\lower0.5ex\hbox{$\sim$}\llap{\raise0.5ex\hbox{$<$}}}\,}

\definecolor{blue}{rgb}{0,0,1}
\definecolor{red}{rgb}{1,0,0}
\definecolor{green}{rgb}{0,1,0}
\newcommand{\blue}[1]{\textcolor{blue}{#1}}
\newcommand{\red}[1]{\textcolor{red}{#1}}
\newcommand{\green}[1]{\textcolor{green}{#1}}
\newcommand{\orange}[1]{\textcolor{orange}{#1}}
\newcommand{\yd}[1]{\textcolor{blue}{#1}}

\newcommand{\scrA}{{\mathcal A}}
\newcommand{\scrE}{{\mathcal E}} 
\newcommand{\scrF}{{\mathcal F}} 
\newcommand{\scrL}{{\mathcal L}}
\newcommand{\scrM}{{\mathcal M}} 
\newcommand{\scrN}{{\mathcal N}}
\newcommand{\scrS}{{\mathcal S}}
\newcommand{\scrs}{{\mathcal s}}
\newcommand{\scrP}{{\mathcal P}}
\newcommand{\scrO}{{\mathcal O}}
\newcommand{\scrR}{{\mathcal R}}
\newcommand{\scrC}{{\mathcal C}}
\newcommand{\scrV}{{\mathcal V}}
\newcommand{\scrD}{{\mathcal D}}
\newcommand{\scrG}{{\mathcal G}}
\newcommand{\scrW}{{\mathcal W}}
\newcommand{\PP}{\mathbb{P}}
\newcommand{\ZZ}{\mathbb{Z}}
\newcommand{\EE}{\mathbb{E}}
\renewcommand{\d}{\mathrm{d}}
\newcommand{\dm}{d_{\rm min}}
\newcommand{\rhojunction}{\rho_{\rm j}}
\newcommand{\rhojunctionLim}{\rho_{{\rm j},0}}
\newcommand{\rhobranch}{\rho_{\rm b}}
\newcommand{\rhobranchLim}{\rho_{{\rm b},0}}
\newcommand{\rhononbridge}{\rho_{\rm n}}
\newcommand{\rhononbridgeLim}{\rho_{{\rm n},0}}
\newcommand{\percolationCluster}{C}
\newcommand{\leafFreeCluster}{C_{\rm \ell f}}
\newcommand{\bridgeFreeCluster}{C_{\rm bf}}
\newcommand{\df}{d_{\rm f}}
\newcommand{\yt}{y_{\rm t}}
\newcommand{\yh}{y_{\rm h}}
\newcommand{\dfprime}{d'_{\rm f}}
\newcommand{\yhhat}{\hat{y}_{\rm h}}
\newcommand{\ythat}{\hat{y}_{\rm t}}
\newcommand{\yhstar}{y^*_{\rm h}}
\newcommand{\ytstar}{y^*_{\rm t}}
\newcommand{\zc}{z_{\rm c}}
\newcommand{\dc}{d_{\rm c}}
\newcommand{\bfx}{{\bf x}}
\newcommand{\bfO}{{\bf o}}
\newcommand{\bfo}{{\bf o}}
\newcommand{\bfS}{{\bf S}}
\newcommand{\bfq}{{\bf q}}
\newcommand{\bfr}{\bf r}
\newcommand{\origin}{\bf 0}
\newcommand{\bfe}{\bf e}
\newcommand{\bfk}{{\bf k}}
\newcommand{\bfy}{\bf y}
\newcommand{\bfu}{\bf u}
\newcommand{\bmomega}{{\bm \omega}}
\newcommand{\bfU}{{\bf u}}
\newcommand{\ind}{\mathbbm{1}}
\newcommand{\xiu}{\xi_{\rm u}}

\newcommand{\Lm}{L_{\rm min}}
\newcommand{\scrT}{{\mathcal T}}

\title{Interplay of the complete-graph and Gaussian fixed-point asymptotics in finite-size scaling of percolation above the upper critical dimension}
\date{\today}
\author{Mingzhong Lu}
\thanks{These two authors contributed equally to this paper.}
\affiliation{Department of Modern Physics, University of Science and Technology of China, Hefei, Anhui 230026, China}
\author{Sheng Fang}
\thanks{These two authors contributed equally to this paper.}
\affiliation{Hefei National Research Center for Physical Sciences at the Microscales, University of Science and Technology of China, Hefei 230026, China}
\author{Zongzheng Zhou}
\email{eric.zhou@monash.edu}
\affiliation{School of Mathematics, Monash University, Clayton, Victoria 3800, Australia}
\author{Youjin Deng}
\email{yjdeng@ustc.edu.cn}
\affiliation{Department of Modern Physics, University of Science and Technology of China, Hefei, Anhui 230026, China}
\affiliation{Hefei National Research Center for Physical Sciences at the Microscales, University of Science and Technology of China, Hefei 230026, China}	
\affiliation{Hefei National Laboratory, University of Science and Technology of China, Hefei 230088, China}	

\begin{abstract}

Percolation has two mean-field theories, the Gaussian fixed point (GFP) and the Landau mean-field theory or the complete graph (CG) asymptotics. By large-scale Monte Carlo simulations, we systematically study the interplay of the GFP and CG effects to the finite-size scaling of percolation above the upper critical dimension $d_c = 6$ with periodic, free, and cylindrical boundary conditions. Our results suggest that, with periodic boundaries, the \emph{unwrapped} correlation length scales as $L^{d/6}$ at the critical point, diverging faster than $L$ above $d_c$. As a consequence, the scaling behaviours of macroscopic quantities with respect to the linear system size $L$ follow the CG asymptotics. The distance-dependent properties, such as the short-distance behaviour of the two-point correlation function and the Fourier transformed quantities with non-zero modes, are still controlled by the GFP. With free boundaries, since the correlation length is cutoff by $L$, the finite-size scaling at the critical point is controlled by the GFP. However, some quantities are observed to exhibit the CG aysmptotics at the low-temperature pseudo-critical point, such as the sizes of the two largest clusters. With cylindrical boundaries, due to the interplay of the GFP and CG effects, the correlation length along the axial direction of the cylinder scales as $\xi_L \sim L^{(d-1)/5}$ within the critical window of size $O(L^{-2(d-1)/5})$, distinct from both periodic and free boundaries. A field-theoretical calculation for deriving the scaling of $\xi_L$ is also presented. Moreover, the one-point surface correlation function along the axial direction of the cylinder is observed to scale as ${\tau}^{(1-d)/2}$ for short distance but then enter a plateau of order $L^{-3(d-1)/5}$ before it decays significantly fast.


\end{abstract}
\pacs{05.50.+q (lattice theory and statistics), 05.70.Jk (critical point phenomena),
64.60.F- (equilibrium properties near critical points, critical exponents)}
\maketitle

\section{Introduction}
\label{Introduction}
Percolation~\cite{stauffer2018introduction} is a fundamental model in statistical mechanics and applied probability, and has wide applications in material science, neuroscience, complex network, epidemiology, etc. Consider the bond percolation model on the hypercubic lattice $\ZZ^d$. Each edge on the lattice is independently occupied by a bond with probability $p$ or empty. Two sites are called connected if there exists a path of connected bonds between them, and a maximal set of connected sites is a \emph{cluster}. A phase transition happens at the critical point $p_c$, that is, for $p < p_c$ all clusters are finite but for $p \geq p_c$ an infinite-large cluster exists. For $d=1$, it is trivial that $p_c = 1$. For $d\geq 2$, it is known that $0 < p_c < 1$. Exact values of $p_c$ are available on many two-dimensional (2D) lattices~\cite{pc_chapter, kesten1980critical}. Only numerical estimates for $p_c$ are available above 2D~\cite{wang2013bond,MertensMoore2018,Grassberger2003}.

The critical behaviour of percolation is characterized by the power-law scaling of various quantities near $p_c$. For example, the order parameter $P_\infty$, defined as the probability that an arbitrarily chosen site belongs to an infinite cluster, scales as $P_\infty \sim (p - p_c)^{\beta}$ as $p$ approaches $p_c$ from above. The correlation length, corresponding to the averaged radius of finite clusters, diverges as $\xi \sim |p - p_c|^{-\nu}$ near $p_c$. The susceptibility $\chi$, which for percolation is the second moment of finite cluster sizes, scales as $\chi \sim |p - p_c|^{-\gamma}$. At $p_c$, the two-point correlation function scales as $g(\bfx) \sim \|\bfx\|^{2-d-\eta}$. The thermodynamic fractal dimension $D_F$ characterizes the scaling of the size of a typical percolation cluster $s$ versus its radius of gyration $R$, i.e., $s\sim R^{D_F}$. Here $\beta, \nu, \gamma, D_F$ and $\eta$ are universal critical exponents, among which only two of them are independent and others are related via the following scaling relations,
\begin{gather}
D_F = d - \beta/\nu\;, \nonumber \\
\gamma = \nu(2 D_F - d)\;, \nonumber \\
\eta = 2 + d - 2 D_F \;. \nonumber
\end{gather}

In 2D, the Coulomb-gas method \cite{nienhuis1982analytical} and the conformal field theory \cite{belavin1984infinite,friedan1984conformal,domb2000phase} predicted that the critical exponents $\beta=5/36$ and $\nu=4/3$. This can even be established rigorously by proving that the scaling limit of percolation is $\rm SLE_6$ (Schramm-Loewner evolution) \cite{lawler2000dimension}. For $2<d<6$, no exact values are available for percolation critical exponents, but high-precision estimates have been obtained using methods like Monte Carlo simulations~\cite{wang2013bond,MertensMoore2018,Grassberger2003}, series expansion~\cite{harris1975renormalization,gracey2015four,janssen1999resistance} and the recently proposed conformal bootstrap method~\cite{leclair2018conformal}. The upper critical dimension of percolation is known to be $d_c=6$~\cite{harris1975renormalization}. For $d\geq d_c$, critical exponents are believed to take mean-field values: $\beta = 1$, $\nu=1/2$, $\gamma = 1$, $D_F = 4$ and $\eta = 0$~\cite{toulouse1974perspectives,priest1976critical}. In sufficiently high dimensions, many rigorous results have been established for percolation~\cite{heydenreich2017progress}. For example, for $d\geq 11$, it has been proved~\cite{FitznerVanderHofstad2017} that $g(\bfx) \sim \|\bfx\|^{2-d}$.

Apart from the thermodynamic-limit (infinite-system) behaviour, the other fundamental problem to study is the \emph{finite-size scaling} (FSS), describing the asymptotic approach of finite-size systems to the thermodynamic limit. The basic assumption in the FSS theory is that the correlation length is cut off by the linear system size $L$, such that, for a typical statistical-mechanical model, the singular part of the free energy density reads
\begin{equation}
\label{Eq:FSS free energy}
    f(t,h) = L^{-d} \tilde{f}(tL^{y_t}, hL^{y_h})\;,
\end{equation}
where the two scaling exponents $y_t, y_h$ are called thermal and magnetic renormalization exponents with $y_t = 1/\nu$ and $y_h = D_F$, and $\tilde{f}$ is the scaling function. The derivatives of $f(t,h)$ with respect to the reduced temperature $t$ and the magnetic field $h$ generate various thermal and magnetic quantities, and thereby their FSS behaviours are also known. In fact, percolation is special since its partition function is constantly 1 and thus the free energy density is zero. But the finite-size behaviour of percolation can still be described by Eq.~\eqref{Eq:FSS free energy}, by studying percolation as the $Q\rightarrow 1$ limit of the general $Q$-state Potts model\cite{wu1982potts}. Below the upper critical dimension $d_c$, FSS has been well understood and proven to be a powerful tool to estimate critical points and exponents. However, FSS above $d_c$ is surprisingly subtle since it depends on the imposed boundary conditions. Taking the Ising model for example, the susceptibility, which corresponds to the second derivative of $f(t,h)$ with respect to $h$, is expected to scale as $\chi \sim L^{2y_h - d}$ at the critical point $t = 0$ and zero field $h=0$. For the Ising model, it is known that $d_c = 4$. Above $d_c$, FSS is expected to be controlled by Gaussian fixed point (GFP) which gives $(y_t, y_h) = (2, 1+d/2)$, and consequently $\chi \sim L^2$. This is true when the system is imposed with free boundary conditions (FBC). However, with periodic boundary conditions (PBC), it was observed that $\chi \sim L^{d/2}$. Many recent works focus on clarifying and deepening the understanding to the FSS above $d_c$; one can refer to the Introduction of Ref.~\cite{fang2021logarithmic} for a brief review. The latest understanding is that, in order to completely describe the FSS of the Ising model with PBC and above $d_c$, one needs two sets of renormalization exponents and the free energy density should be written as
\begin{equation}
\label{eq:free_energy}
f(t,h) = L^{-d}\tilde{f}_0(tL^{y_t^{\ }},hL^{y_h^{\ }}) + L^{-d} \tilde{f}_1(tL^{y^*_t}, h L^{y^*_h})\;.
\end{equation}
Apart from the GFP exponents $(y_t, y_h)$, it involves another set of exponents $(y^*_t, y^*_h) = (d/2, 3d/4)$, which are inferred from the Ising model on the complete graph (CG). FSS of various quantities can be understood as the interplay of the CG asymptotics and the GFP predictions; the former describes the FSS of macroscopic quantities while the latter controls spatial fluctuations and short-distance behaviour. For example, the susceptibility follows the CG-asymptotics $\chi \sim L^{2y^*_h -d } = L^{d/2}$. The correlation function $g(\bfx, L) \approx \|\bfx\|^{2-d} + L^{-d/2}$, namely the short-distance behaviour is controlled by GFP but for larger distance it is dominated by CG-asymptotics. A systematic study of how this interplay affects the FSS of various quantities of the Ising model with PBC and $d\geq d_c$ can be found in Refs.~\cite{Lv2020Two,FangZhouDeng2023,Fang2022Geometric,fang2020complete}.

Percolation also has two sets of renormalization exponents; one is $(y_t, y_h) = (2, 1+d/2)$ from the GFP which is the same as the Ising model, and the other one is $(y^*_t, y^*_h) = (d/3, 2d/3)$ inferred from percolation on the complete graph. Therefore, it is natural and interesting to ask whether one can observe the interplay of the CG and the GFP asymptotics in the FSS of various quantities of percolation, and moreover, how the effect of interplay depends on boundary conditions. To investigate this question, we study in this paper the bond percolation model on 7D hypercubic lattices with PBC, FBC and cylindrical boundary conditions (CBC).

We start with discussing the PBC case. In Ref.~\cite{huang2018critical}, it was observed that at $p_c$ the size of the largest cluster $C_1 \sim L^{2d/3}$ at 7D, i.e., the finite-size fractal dimension $D_{L} = 2d/3$ which is equal to $y^*_{h}$. This scaling of $C_1$ can be rigorously proved when $d$ is sufficiently large~\cite{HeydenreichVanderHofstad2007,heydenreich2017progress}. On the CG with volume $V$, it is known that $C_1\sim V^{2/3}$, i.e., the finite-volume fractal dimension $D_{V} = 2/3$. Thus, one can safely conjecture that $C_1$ follows the CG asymptotics for all $d\geq d_c = 6$. The second moment of sizes of all clusters is the susceptibility $\chi$, and our data suggest that $\chi \sim L^{2y^*_{h} -d} = L^{d/3}$ which again follows the CG asymptotics. Apart from the sizes, one can also study the cluster-number density function $n(s,L)$, which is defined by interpreting $L^d n(s,L){\rm d}s$ as the number of clusters with size in the range $[s, s+{\rm d}s)$. It is expected that $n(s,L) \sim s^{-\tau} \tilde{n}(s/L^{D_L})$, where $\tilde{n}(x)$ is a universal function which decays exponentially fast when $x \gg 1$. By the standard scaling relation, the Fisher exponent $\tau = 1 + d/D_L$ , which is equal to $5/2$ for percolation above $d_c$. The value $\tau = 5/2$ has been numerically observed for percolation at 7D and also on the CG~\cite{huang2018critical}. Thus, these distance-independent quantities, such as $C_1, \chi, n(s,L)$ follow the CG asymptotics.

However, the two-point correlation function $g(\bfx, L)$, which characterizes the spatial fluctuations, has been shown to exhibit the following behaviour in sufficiently high dimensions~\cite{hutchcroft2021high},
\begin{equation}
\label{eq:correlation_function_finite}
g({\bf x}, L) \approx c_1 \|{\bf x}\|^{2-d} + c_2 L^{-2d/3}\;,
\end{equation}
with some constants $c_1, c_2$. Namely, the distance-dependent part is controlled by the GFP and thus the exponent $\eta = 0$. Moreover there exists a background term $L^{-2d/3}$ which contributes to the leading scaling of the susceptibility since $\chi = \sum_{\bfx} g(\bfx, L)$ and thus follows the CG asymptotics. The crossover happens at $\|\bfx\| \sim L^{2d/[3(d-2)]}$. So one can clearly see the interplay of the CG and GFP effects in the behaviour of $g(\bfx, L)$. Since the background term is distance-independent, it can be removed by studying the Fourier transformation of $g(\bfx, L)$, denoted by $\chi_{\bfk}$. Indeed, our data clearly show that $\chi_{\bfk} \sim L^{2y_h -d} = L^2$ for non-zero modes $\bfk$, where $y_h = 1 + d/2$ from the GFP. Therefore, for percolation above $d_c$, although many macroscopic quantities follow the CG asymptotics, the spatial fluctuations are still controlled by the GFP.

A geometric way to effectively recover the GFP behaviour is to consider an appropriate length scale which correctly accounts for the effect of periodic boundaries. Instead of the system size $L$, one can use the \emph{unwrapped} radius of gyration $R$ for the characteristic length of percolation clusters, which is effectively the radius of gyration if the clusters on the torus are embedded onto the infinite lattice. Indeed, our data show that the scaling $C \sim R^{D_F}$ holds for all clusters, with $D_F = 4$ from the GFP predictions. Moreover, one can define the unwrapped correlation length $\xi_u$ of the system as $R_1$, the unwrapped radius of the largest cluster. Combining the scaling $C_1 \sim L^{D_L}$ and $C_1 \sim R^{D_F}_1$, it immediately follows that $\xi_u, R_1 \sim L^{D_L/D_F} =  L^{d/6}$. Namely, unlike the traditional correlation length which has to be cut off by $L$ in finite systems, the unwrapped correlation length diverges much faster than $L$ if $d > 6$. In other words, the largest cluster winds around the torus extensively. This motivates defining a new topological exponent $\kappa$ to characterize the winding phenomena of percolation clusters above $d_c$, and $\kappa = D_{L}/D_F$ connecting the finite-size and thermodynamic fractal dimensions. For susceptibility, one can see that $\chi \sim L^{d/3} \sim \xi^2_u$, which is consistent with the GFP prediction $\chi \sim L^2$ but with $L$ replaced by $\xi_u$. One can also expect that, if the two-point function is defined in the unwrapped length scale, then it will simply follow the GFP prediction, i.e., without the background term in Eq.~\eqref{eq:correlation_function_finite}. But, for some other quantities which are independent on length scale, such as $n(s,L)$, their behaviour remain the same under the unwrapped length scale.

The other quantity, from which one can see the interplay of the CG and GFP effects, is the number of spanning clusters $N_s$, defined as the number of clusters with unwrapped radii no less than $L$. A cluster with $R \geq L$ has size at least of order $L^4$, where the exponent $4$ is from the GFP. It follows that $N_s = \int_{L^4} L^d n(s,L) {\rm d}s \sim L^{d-6}$, where we use the CG-value $\tau = 5/2$ in $n(s,L)$. The scaling of $N_s$ is confirmed by our numerical results.

We next discuss the FBC case. Since now the correlation length is cut off by $L$, the CG asymptotics is absent and the critical behaviour simply follows the GFP prediction. Consequently, at the critical point, the two-point function $g(\bfx, L) \sim \|\bfx\|^{2-d}$, and the susceptibility $\chi \sim L^2$. The size of the largest cluster scales as $C_1 \sim L^{D_F} = L^4$~\cite{aizenman1997number}, and the cluster-number density $n(s,L)\sim s^{-5/2}\tilde{n}(s/L^4)$. The number of spanning clusters can be similarly obtained as $N_s \sim L^{d-6}$. We note that the common relation $D_F = y_h$ fails for the FBC case above 6D, and this can be explained by the divergence of $N_s$. The total contribution of spanning clusters to the susceptibility $\chi$ is $L^{-d} N_s L^{2 D_F} \sim L^{2D_F -6}$. Combining with $\chi \sim L^{2y_h - d}$, we have $D_F = y_h - d/2 + 3$. It follows that $D_F = y_h$ at 6D but $D_F < y_h$ above 6D.

For FBC, another interesting question to investigate is whether one can observe some PBC-like behaviours at some pseudo-critical point $p_L$. One way to define $p_L$ is as follows. In simulations, we start with the empty lattice and then uniformly at random place bonds one by one onto the lattice. Following \cite{nagler2011impact}, we define $p_L$ as the bond density at which the size of the largest cluster has the maximum rate of increase. Obviously, $p_L$ is random and we denote $\left\langle p_L\right\rangle$ its expectation. At 7D, our data suggest that $\left\langle p_L\right\rangle - p_c \sim L^{-2}$, and it is in the low-temperature regime. More interestingly, at $p_L$, our data clearly show that $C_1 \sim L^{2d/3} = L^{y^*_h}$, the same scaling as the PBC case at $p_c$. Moreover, we observe that the standard deviation of $p_L$ is of order $L^{-y^*_t} = L^{-d/3}$, which is also the same as the critical-window width of the PBC case. Clearly, since $L^{-d/3}$ decays faster than $L^{-2}$ when $d > 6$, we know $p_c$ is out of the standard deviation range of $p_L$, consistent with the fact that the FBC case at $p_c$ shows different critical behaviour as at $p_L$. Note that, for FBC at $p_L$, not all quantities exhibit the PBC behaviour. For example, the correlation length cannot exceed $L$, and the behaviour of the two-point function is still unclear. Whether $g(\bfx)$ exhibits the PBC behaviour or it is power-law but with a new exponent needs further investigation.

We finally study the CBC case at 7D, which is constructed by letting one dimension ($\tau$-direction) be half-infinite ($\tau \in [0,\infty)$) and other 6 dimensions periodic with the linear size $L$. It is interesting to investigate the effect of the interplay from the $(d-1)$-dimensional tori and the 1D infinite system on the finite-size scaling of the CBC case. Initially ($\tau=0$), all edges on the boundary with volume $L^{d-1}$ are occupied, and then a cluster is growing towards the positive $\tau$ direction. We study the correlation length $\xi_L$, which measures how far the cluster grows to in the $\tau$ direction. By a field-theoretical calculation, we obtain that $\xi_L \sim L^{(d-1)/5} \widetilde{\xi}_L\left[ t L^{2(d-1)/5} \right]$ with $t \propto (p - p_c)$, and confirm this scaling numerically. Thus, in the CBC case, the interplay effect produces a new scaling for the correlation length, distinct with both the PBC and the FBC cases. Moreover, we observe the one-point surface correlation function $g_s(\tau) \sim \tau^{-(d-1)/2}$ when $\tau\leq O(\xi_L)$, and then enters a plateau of order $L^{-3(d-1)/5}$ before decaying significantly fast.


The remainder of the paper is organized as follows. In Section~\ref{Sec:picture}, we present a physical picture to understand the FSS of percolation with the three studied boundary conditions above $d_c$. In Sec.~\ref{Sec:field-theory}, we provide an understanding from the field theory to the interplay of the CG and GFP effects, and a field-theoretical calculation for deriving the scaling of the correlation length of the CBC case. In Section~\ref{Sec:Simulations}, details for simulations and observables are summarized. Data analysis to various quantities are presented in Section~\ref{Sec:Results}. Finally, we conclude this paper in Section~\ref{Sec:Discussion}.

\section{Physical picture for the three boundary conditions}
\label{Sec:picture}
For $d\geq d_c$, the critical behaviour of percolation on the infinite lattice is controlled by the GFP. At $p_c$, the two-point function $g(\bfx) \sim \|\bfx\|^{2-d}$ and the correlation length $\xi$ diverges. When the free boundary conditions are imposed, obviously the correlation length $\xi_L$ is cutoff at $L$. This means that $g(\bfx, L)$ still follows the GFP prediction, but the power-law behaviour can only extend to $L$. Consequently, one has the susceptibility $\chi := \sum_{\bfx}g(\bfx, L) \sim L^2$. So, the finite-size scaling of the FBC case still follows the GFP prediction.

Given a box with FBC, if the two sides in each dimension of the box are connected, then the box is with periodic boundary conditions (torus). Thus, on torus, percolation clusters can cross the boundary and wrap around the box. In low dimensions ($d < d_c$), a percolation cluster only winds around the box of constant times, so percolation with PBC exhibits the same finite-size scaling as with FBC. However, in high dimensions ($d > d_c$), the number of times percolation clusters wind around the box diverges with $L$. Such a proliferation of winding acts as the background (no spatial fluctuations) which results that macroscopic quantities follow the CG asymptotics, for example, the sizes of large percolation clusters scale as $V^{2/3}$ both on CG and tori with volume $V = L^d$ and $d\geq d_c$. The CG has no spatial fluctuations, and moreover it has large vertex degree and is translational invariant, just like the high-dimensional torus. The effect of winding on the two-point function is that, in addition to the power-law behaviour, a distance-independent term $L^{-2d/3}$ is added to $g(\bfx, L)$, and the term $L^{-2d/3}$ corresponds to the correlation between any two vertices on the CG. Thus, for short distance, $\|\bfx\|^{2-d}$ dominates $L^{-2d/3}$, so $g(\bfx, L)$ still follows the GFP prediction. While for large distance, $g(\bfx, L)$ enters the distance-independent plateau $L^{-2d/3}$. Consequently, since $\chi := \sum_{\bfx}g(\bfx, L)$, one has $\chi \sim L^{d/3}$ which also follows the CG asymptotics. The prediction from GFP, which is $L^2$, becomes subdominant in $\chi$. Since the CG asymptotics on tori is due to the proliferation of winding, one can expect to recover the GFP behaviour from it by defining a proper length scale which correctly accounts for the effect of winding. A natural choice is to use \emph{unwrapped} length scale to replace the standard Euclidean distance. Indeed, our data show that, in terms of the \emph{unwrapped} length scale, quantities like the sizes of percolation clusters and two-point function follow the GFP predictions.


The above argument can be made more explicitly from the aspect of scaling hypothesis. Let $t$ denote the deviation from the criticality. Consider a generic quantity $\scrO$ which scales as $\scrO \sim |t|^{-y_\scrO}$ near the critical point. Since the correlation length diverges as $\xi \sim |t|^{-\nu}$, it follows that $\scrO \sim \xi^{y_\scrO/\nu}$. Standard scaling hypothesis further conjectures that it can be written as $\scrO \sim \xi^{y_\scrO/\nu} \widetilde{\scrO}(t\xi^{1/\nu})$ with $\widetilde{\scrO}(\cdot)$ the scaling function. For PBC, the unwrapped correlation length diverges as $\xi_L \sim L^{d/6}$ for $d \geq 6$. If one uses $\xi_L$ instead of $L$ in the scaling formula, then
\begin{equation}
\label{Eq:PBCtoGFP}
\scrO \sim \xi^{y_\scrO/\nu}_L \widetilde{\scrO}(t \xi^{1/\nu}_L) \sim L^{\frac{d}{3}y_\scrO} \widetilde{\scrO}(tL^{d/3})\;, 
\end{equation}
where $\nu = 1/2$ is used. The right-hand-side of Eq.~\eqref{Eq:PBCtoGFP} is exactly the scaling formula for quantities in the PBC case. Taking the susceptibility $\chi$ for example where $y_\scrO = \gamma = 1$, one has $\chi \sim L^{d/3}\widetilde{\chi}(tL^{d/3})$.

A long-standing interesting question is that whether one can observe the critical FBC behaviour in the PBC case, and vice versa. From the above argument, one can expect that for the PBC case, if at some $p < p_c$ such that the unwrapped correlation length decreases to the order of $L$, then the critical FBC behaviour can be observed. We believe this happens at $p_L = p_c - cL^{-2}$ with arbitrary constant $c>0$, i.e., in the high-temperature scaling window. Conversely, in the FBC case, our data suggest that at the pseudo-critical point $p_L$, where $p_L = p_c + cL^{-2}$ with some constant $c>0$ (low-temperature scaling window), the PBC behaviour at criticality can be observed.

For CBC, by a field-theoretical analysis, we show that the correlation length in the axial direction is $\xi_L \sim L^{(d-1)/5} \widetilde{\xi}_L(tL^{2(d-1)/5})$, which at $p_c$ is greater than both $L$ and $L^{d/6}$ for $d > 6$. So, for CBC, when a percolation cluster is growing from the origin, the effective volume it can explore is $V_{\rm eff} = \xi_L L^{d-1} \sim L^{6(d-1)/5}$. In terms of $V_{\rm eff}$, one can rewrite $\xi_L \sim V^{\frac{1}{6}}_{\rm eff} \widetilde{\xi}_L(tV^{\frac{1}{3}}_{\rm eff})$. This is the same scaling as the unwrapped correlation length in the PBC case, where the effective volume is just $V = L^d$, and the exponent $\frac{1}{3}$ is the value of $\ytstar$ from the complete-graph. Therefore, we conjecture that the CBC system with linear size $L$ exhibits the similar behaviour as the PBC system with volume $V_{\rm eff}$. Furthermore, same to the PBC case, we conjecture that for the CBC case, if one uses the unwrapped length scale, then the GFP behaviour can also be recovered.

To sum up, we argue that the distinct critical behaviours observed on the high-dimensional boxes with FBC, PBC, and CBC can be reunified by replacing the standard Euclidean length scale with the unwrapped length scale which can correctly account for the boundary effects.

\section{Understanding from field theory}
\label{Sec:field-theory}
\subsection{The PBC case}
\label{Sec:PBC-field-theory}
Field theory provides a powerful framework for predicting universal critical behaviours of many-body systems, such as determining critical exponents and scaling laws. We elaborate in this section the understanding to the FSS of the percolation model above $d_c$ from the field theory, which is consistent with the physical picture in Section~\ref{Sec:picture}. We start with the standard $\phi^3$ theory which can be applied to describe the critical behaviour of percolation~\cite{houghton1978high,potts1952some}. 

In the $\phi^3$ field theory, the action of a $d$-dimensional system in real space is
\begin{equation}
    S[\Phi] = \int \mathrm{d}^d x \left\{\frac{1}{2} (\nabla \Phi)^2 + \frac{1}{2} r_0 \Phi^2 + \frac{1}{6} w \Phi^3 \right\},
    \label{eq:phi_3_field}
\end{equation}
where $\Phi(\bfx)$ is a field related to the order parameter of the system, $r_0$ is the coefficient of the quadratic term which relates to the temperature (for percolation, it is the bond occupation probability), and $w$ is a parameter~\cite{zia1975critical,amit1977universality}. It is conjectured that near the critical point $r_0$ is linear in the reduced temperature $t$. By applying dimensional analysis and noting that the action $S$ has scaling dimension 0, we deduce that the scaling dimensions of $\Phi$, $r_0$ and $w$ are $\frac{d}{2}-1$, $y_t = 2$ and $y_w = 3-\frac{d}{2}$, respectively. Apparently, when $d > d_c = 6$, one has $y_w<0$ and the term $w\Phi^3$ becomes irrelevant. In the parameter space $(t, w)$, the Gaussian fixed point is $(t, w) = (0,0)$.

Usually, it is more convenient to study the action $S[\Phi]$ in the momentum space, i.e., using Fourier-transformed $\Phi$. But the Fourier modes of $\Phi(\bfx)$ have different forms under different boundary conditions. For PBC, we can expand $\Phi(\bfx)$ in Fourier modes as
\begin{equation}
    \Phi\left(\bfx\right)=\sum_{\bfk} \mathrm{e}^{i \bfk \cdot \bfx} \varphi_{\bfk},
    \label{eq:PBC_Fourier_ex}
\end{equation}
in which the components of $\bfk$ are discrete, in units of $2 \pi / L$. It is clear that only the non-zero modes $\varphi_{\bfk\neq 0}$ contain spatial fluctuations. Applying $r_0 = a t$ with some constant $a$ near the critical point, we can separate the action as
\begin{equation}
    S[\Phi] = L^d \left\{\frac{a}{2} t \varphi^2_0 + \frac{1}{6} w \varphi^3_0 \right\} + S_{\bfk \neq 0},
\end{equation}
and we call the first term in the RHS as $S_{\bfk = 0}$ which is effectively the CG percolation, and, for $d > d_c = 6$, the second term corresponds to the GFP. We next discuss the contribution of zero and non-zero modes to various quantities for $d > d_c = 6$.

First, zero and non-zero modes contribute the same order to the action $S$, both of the volume order. Second, all terms contribute to the location of the critical point. The critical point is $t=0$ if all non-zero modes are neglected. Including non-zero modes to the action will shift the critical point to a non-zero value. But such a shift tends to vanish as $d$ goes to infinity. Third, if one studies the leading scaling behaviour of macroscopic quantities, it suffices to include the zero-mode term and the terms with non-zero modes only provide subdominant contribution. For example, using only the zero mode, one can obtain the susceptibility $\chi \sim L^{d/3}$, consistent with the CG asymptotics. The non-zero modes, which control spatial fluctuations, contribute a term of order $L^2$ which is sub-dominant and corresponds to the GFP predictions. However, if we study the two-point function, then its behaviour (excluding the background term $L^{-2d/3}$) is controlled by the non-zero modes, i.e., $g(\bfx, L)\sim \|\bfx\|^{2-d}$ (or $\eta = 0$) from the GFP. Therefore, from field theory, the interplay from zero and non-zero modes gives the same picture as the interplay of the CG and GFP effects, as discussed in Sec.~\ref{Sec:picture}.

\subsection{The CBC case}
\label{Sec:CBC-field-theory}
We present in this section a field-theory analysis to derive the scaling of the correlation length in the CBC case. In \cite{brezin1985finite}, the authors derived the scaling of the correlation length for systems described by the $\phi^4$ field theory with CBC. Here we extend their method to the $\phi^3$ field theory. For systems with CBC, one can write the field $\Phi(\bfx) = \Phi(\boldsymbol{x}, \tau)$ where $\boldsymbol{x}$ and $\tau$ denote the coordinates in the $d-1$ periodic directions and the half-infinite $\tau$ direction, respectively. The field $\Phi(\bfx)$ is expanded in Fourier modes only in the periodic $(d-1)$ directions,
\begin{equation}
    \Phi\left(\boldsymbol{x}, \tau\right)=\sum_{\bfq} \mathrm{e}^{i \bfq \cdot \boldsymbol{x}} \varphi_{\bfq}(\tau),
    \label{eq:CBC_Fourier_ex}
\end{equation}
in which the components of $\bfq$ are discrete, in units of $2 \pi / L$. Substituting Eq.~\eqref{eq:CBC_Fourier_ex} into Eq.~\eqref{eq:phi_3_field} gives the expression for the action
\begin{equation}
\begin{split}
    S = L^{d-1}\int &\mathrm{d}\tau \biggl\{ \frac{1}{2}\dot{\varphi}^2 + \frac{a t}{2}\varphi^2 + \frac{w}{6} \varphi^3 \biggl\} + S_{\boldsymbol{q} \neq 0},
\end{split}
\label{eq:action_all_q}
\end{equation}
where $\varphi \equiv \varphi_{\bfq=0}(\tau)$. All terms with $\bfq \neq 0$ are absorbed into $S_{\bfq \neq 0}$. By the same argument as the PBC case, the non-zero modes contribute subdominantly to the scaling of the correlation length, and thus can be neglected. Denote the first term in the RHS of Eq.~\eqref{eq:action_all_q} as $S_{\mathrm{eff}}[\varphi]$, then the partition function can be written as
\begin{equation}
    Z = \int \mathrm{D}[\varphi] \exp(-S_{\mathrm{eff}}[\varphi])\;,
\end{equation}
where $\int D[\varphi]$ denotes the functional integral.
This coincides with the Feynman path integral of a single-particle quantum mechanical problem along the imaginary time direction, and
it can be seen more clearly by writing down the propagator
\begin{equation}
    \langle q_f|e^{-\hat{H}T}|q_i \rangle = \int_{q(0)=q_i}^{q(T)=q_f} \mathrm{D}[q] 
 \exp\left[-\int_0^T \mathrm{d} \tau  H(q,\dot{q})\right],
\end{equation}
This is, starting at the initial state $q_i$, the probability of arriving at the final state $q_f$ in time $T$.
Here $H(q,\dot{q}) = m\dot{q}^2/2+V(q)$ is the Hamiltonian where $\dot{q}$ is velocity and $V(q)$ is the potential.
Comparing the integrand of $S_{\mathrm{eff}}[\varphi])$ with $H(q,\dot{q})$, one can notice that $\varphi$ corresponds to $q$, $L^{d-1}$ corresponds to the mass $m$ and the momentum $p=m\dot{q}$ corresponds to $L^{d-1} \dot{\varphi}$. Thus, the Hamiltonian corresponding to the action in Eq.~\eqref{eq:action_all_q} is
\begin{equation}
    \hat{H}(p, q)=\frac{\hat{p}^{2}}{2 L^{d-1}}+L^{d-1}\left[\frac{at}{2} q^{2}+\frac{w}{6}q^{3}\right].
    \label{eq:hami}
\end{equation}

We now study the energy eigenvalues of the Hamiltonian operator $\hat{H}$. The explicit eigenvalues are not easy to find, but the scaling behaviour of the eigenvalues can be extracted by properly rescaling $q$ and $\hat{p}$. We first dilate $q$ to $q^{\prime} := L^{2(d-1)/5} w^{1/5} q$, and consequently, since $\hat{p} = -i\frac{\partial}{\partial q}$, $\hat{p}$ is rescaled to $\hat{p}' = L^{-2(d-1)/5} w^{-1/5} p$. Then we can write
\begin{equation}
    \hat{H} =w^{2/5} L^{-(d-1)/5}\left(\frac{1}{2} \hat{p}^{\prime2}+ \frac{1}{2} t_r q^{\prime2} + \frac{1}{6}q^{\prime3}\right), \nonumber
\end{equation}
with $t_r = a t w^{-4/5} L^{2(d-1)/5}$. Therefore we can write the set of energy eigenvalues as
\begin{equation}
    E_{\alpha}\left(t, L\right) = L^{-(d-1) / 5} \widetilde{E}_{\alpha}\left[t L^{2(d-1) / 5}\right], \nonumber
\end{equation}
where $\alpha \in \{0,1,2,\cdots\}$ and $\widetilde{E}_{\alpha}$ denotes the scaling function. In quantum mechanics, the inverse of the gap ($E_1 - E_0$) between the lowest two energy levels is related to the correlation length~\cite{zinn2021quantum}, and thus we have
\begin{equation}
    \xi_{L}(t) = L^{(d-1) / 5} \tilde{\xi}_L \left[t L^{2(d-1) / 5} \right],
    \label{eq:corre}
\end{equation}
with some scaling function $\tilde{\xi}_L(\cdot)$. Now we can see that the way we dilate $q$ to $q^\prime$ in Eq.~\eqref{eq:hami} is to ensure $\xi_L$ can be written in this standard FSS formula, which implies that $\xi_{L} \sim L^{(d-1)/5}$ at the critical point $t=0$ and also within a critical window with size $O(L^{-2(d-1) / 5})$. Thus, in contrast to the PBC and FBC cases, the interplay of the CG and GFP effects in the CBC case produces a new scaling behaviour for the correlation length.

\section{SIMULATION and OBSERVABLE}
\label{Sec:Simulations}

\subsection{7D percolation with PBC}
\label{Sec:Simu-PBC}
We simulate the 7D bond percolation model with PBC at the critical point $p_c = 0.078\,675\,230(2)$~\cite{mertens2018percolation} with even $L$ from 4 to 28. Since $p_c$ is small, we apply the trick used in~\cite{henk2002cluster,deng2005monte} to speed up the bond placing process. After a bond configuration is generated, we use the breadth-first search to identify and measure all clusters. In order to measure the Fourier-transformed susceptibility, we artificially define a spin variable $\sigma_\scrC$ for each cluster $\scrC$, where $\sigma_\scrC$ takes either 1 or $-1$ uniformly at random:

In simulation, for each bond configuration we sample the following  quantities.
    \begin{enumerate}
        \item  The size of the largest cluster $\mathcal{C}_1$;
        \item The pseudo-magnetization  $\mathcal{M} = \sum_{\{\scrC\}} |\scrC| \sigma_{\scrC}$ where $|\scrC|$ denotes the size (number of vertices) of the cluster $\scrC$;
        \item The Fourier mode of the pseudo-magnetization $ \mathcal{M}({\bf{k}}) =  \sum_{\{\scrC\}} \sigma_{\scrC} \sum_{j \in \scrC} e^{{i {\bf k} \cdot {\bf r}_j}} $, in which the wave vector ${\bf{k}} = \frac{2\pi}{L}(1,0,0,0,0,0,0)$ and $\bf{r_j}$ is the Euclidean coordinate of site $j$;
        \item The \emph{unwrapped} relative coordinate $\bf{u}$ of every vertex in each cluster. For a cluster $\scrC$, we started with the vertex $o$, which has the smallest label according to some arbitrary but fixed vertex labelling, and set ${\bf{u}}_o = 0$. We breadth first search through the cluster. For a newly visited site $v$, we set ${\bf{u}}_{v} = {\bf{u}}_{u} + \bf{e}_i (- \bf{e}_i)$ if the vertex $v$ is traversed from $u$ along (against) the $i$th direction, where $\bf{e}_i$ is the unit vector in the $i$th direction;
        \item The unwrapped extension for each cluster, which is defined as $\mathcal{U} (\scrC) = \max_{\bfx,\bfy \in \scrC}( | {\bf{u}}_{\bfx}^{1} -{\bf{u}}_{\bfy}^{1} | )$. Here ${\bf{u}}_{x}^{1}$ denotes the first entry of ${\bf{u}}_{x}$;
        \item The indicator function $\mathcal{P}_{L}(\cdot)$ for each cluster, defined as
                   \begin{equation}
                     \mathcal{P}_{L}(\scrC) = 
                        \left \{
                        \begin{array}{lcl}
                            1    &     & \text{if $\mathcal{U}(\scrC) \ge L$} \\
                            0    &     & \text{otherwise} \ ;\\
                        \end{array}
                         \right.  \nonumber
                    \end{equation} 
        \item The unwrapped radius of gyration $\scrR(\cdot)$ for each cluster, defined as 
          \begin{align}
             \mathcal{R}(\scrC) = \sqrt{ \sum_{\bfx\in \scrC} \frac{( {\bf{u}}_{\bfx} - \bar{ {\bf{u}}} )^2}{|\scrC|} } \;,
              \nonumber
          \end{align}
      with $\bar{{\bf{u}}} = \sum_{\bfx \in \scrC} { \bf{u} }_{\bfx}/ |\scrC|$.
        \end{enumerate}
      We then calculate the following quantities.
        \begin{enumerate}
            \item The mean size of the largest cluster $C_1  = \langle \mathcal{C}_1 \rangle$; 
            \item The susceptibility $\chi = L^{-d} \langle \sum_{\scrC} |\scrC|^2 \rangle$ and its Fourier mode $\chi_{\bf{k}} = L^{-d} \langle | \mathcal{M}(\bf{k}) |^2  \rangle $;
            \item The mean radius of gyration of the largest cluster $R_1$;
            \item $R(s)$, the averaged radius of gyration from clusters with sizes in $[s, s+\Delta s)$;
            \item The number of spanning cluster $N_{\rm s,0} = \langle \sum_{\scrC} \mathcal{P}_{L}(\scrC)  \rangle $ and $N_{\rm s,1} = \langle \sum_{\scrC} \mathcal{P}_{L-1}(\scrC)  \rangle $.
        \end{enumerate}

       \begin{figure}[t]
            \centering
            \includegraphics[scale=0.5]{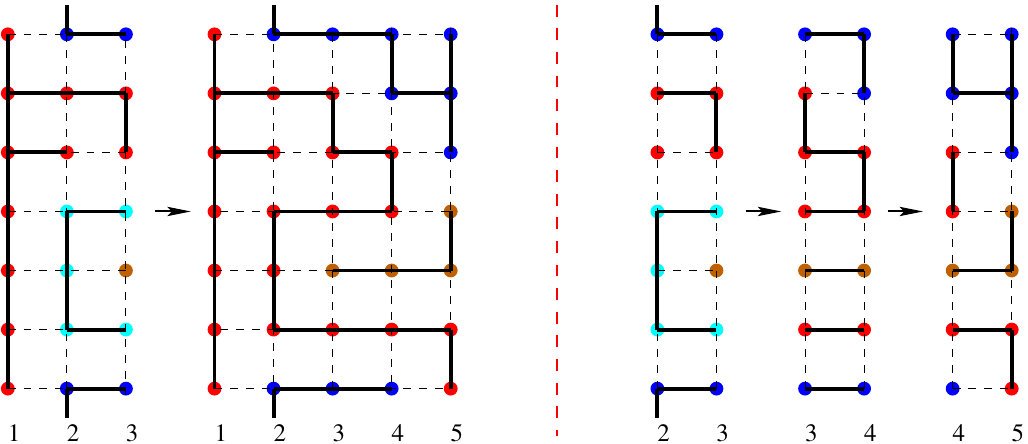}
            \caption{An example bond configuration on a two-dimensional cylinder. The left two figures show that one can read the surface connectivity on layer $\tau$ if all bonds information up to layer $\tau$ is recorded. For instance, on layer 3, there are 4 clusters, and 2 sites (red) belong to the same cluster as the initial layer. On layer 5, there are 3 clusters and 2 sites belong to red cluster. The right three figures show that, using TMC, one only needs to record the bond information on consecutive two layers. It can be seen that the surface connectivity at layers 3 and 5 are the same as the left figures.}
            \label{fig:2D}
        \end{figure}

\subsection{7D percolation with FBC}
\label{7DFBC}
We also simulate bond percolation in seven dimensions with FBC. We randomly place bonds to the lattice one by one and then measure the size of the largest cluster $\scrC_1(n)$ and its increase rate $\Delta(n) = \scrC_1(n+1) - \scrC_1(n)$, as a function of the number of bonds $n$ on the lattice. Following \cite{nagler2011impact}, we define the bond density, at which $\Delta(n)$ reaches its maximum value, as the pseudo-critical point $\tilde{p}_L$. In each simulation, we first locate $\tilde{p}_L$, and then at $\tilde{p}_L$ we sample the size of the largest cluster $\scrC_1$.
By taking the ensemble average, we then calculate the mean value of the pseudo-critical point $p_L = \langle \tilde{p}_L \rangle$, its fluctuation $\sigma_{p} = \sqrt{\left<(\tilde{p}_L-\langle \tilde{p}_L \rangle)^2\right>}$, and the mean value of the largest cluster $C_1  = \langle \mathcal{C}_1 \rangle$.

The system sizes being simulated are even values from 4 to 16. About $2.5 \times 10^5$ samples are generated for each $L < 12$ and $7.2 \times 10^4$ samples for each $L \geq 12$.

       \begin{figure}[t]
            \centering
            \includegraphics[scale=0.67]{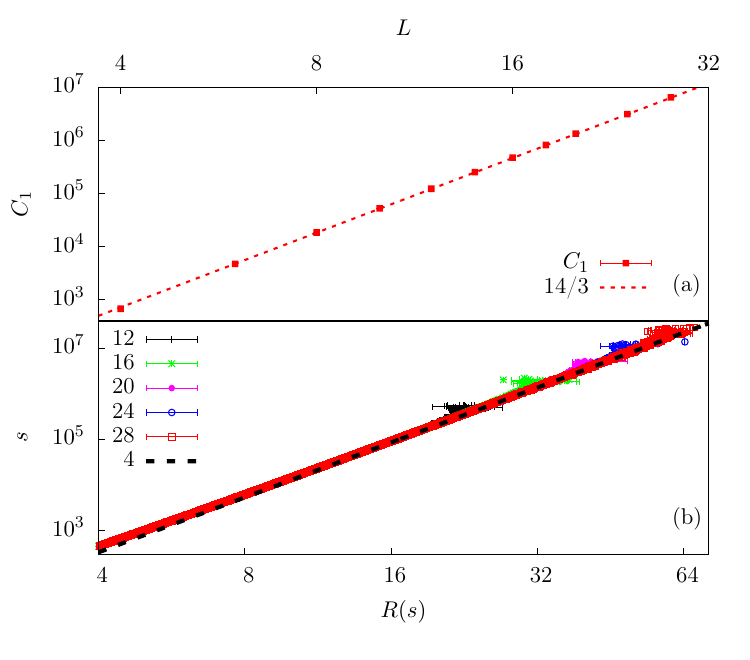}
            \caption{Plot to demonstrate the finite-size and thermodynamic fractal dimensions. (a) Log-log plot of the size of the largest cluster $C_1$ versus the system size $L$, which suggests it scales as $C_1 \sim L^{14/3}$; (b) Log-log scatter plot of a cluster's size $s$ versus its radius of gyration $R(s)$ for all percolation clusters, it implies that $R \sim s^4$ which is controlled by the GFP. }
            \label{fig:C1_ckra}
        \end{figure}

\subsection{7D percolation with CBC}
\label{7DCBC}

We use the transfer Monte Carlo (TMC) algorithm~\cite{deng2005monte} to simulate the bond percolation model on the $L^6 \times \infty$ cylindrical lattices with even $L$ from 4 to 24. We start with a torus with volume $L^{d-1}$, as the initial layer ($\tau = 0$), where all edges are occupied. We then add a new layer ($\tau = 1$) of torus of the same volume next to the initial layer, and then uniformly at random with probability $p_c$ occupy the edges on layer $1$ and the edges connecting layer 0 and layer 1. We record the cluster information at $\tau = 1$, i.e., which sites at $\tau = 1$ belong to the same cluster. Note that, if two sites at layer $\tau$ belong to the same cluster, they can be connected either through a path on the layer $\tau$ or a path going back through previous layers, or both. The information after layer $\tau$ has no effect to the connectivity at layer $\tau$, and thus what we study here is the surface property. At each layer $\tau$, we sample $\scrN(\tau)$, the number of sites belong to the same cluster as sites on the initial layer, and the one-point surface correlation function $g_s(\tau) = \langle\scrN(\tau)\rangle/L^{d-1}$. Let $\tau_{\rm max} := \max\{\tau: \scrN(\tau) > 0\}$, then the ensemble average of $\tau_{\rm max}$ measures the correlation length $\xi_L$ along the $\tau$-direction. The sketch in Fig.~\ref{fig:2D} demonstrates the simulation process for a particular configuration using the TMC algorithm. Using this algorithm, we only record the cluster information at the current layer and the forthcoming layer. It saves memory space and suffices to measure the correlation length and the one-point surface correlation function.

\section{Results}
\label{Sec:Results}
We perform in this section a detailed data analysis to various quantities. To extract the leading scaling behaviour of a generic observable $\scrO$ at the critical point, we perform the least squares fitting of the $\scrO$ data to the ansatz 
\begin{equation}
\label{eq:fitting_ansatz}
\scrO = L^{y_{\scrO}} (a_0 + a_1 L^{y_1} + a_2L^{y_2}) +c_0,
\end{equation}
where $y_1, y_2 < 0$ are correction-to-scaling exponents. As a precaution against correction-to-scaling terms that we miss including in the fitting ansatz, we impose a lower cutoff $L \ge \Lm$ on the data points admitted in the fit and systematically study the effect on the residuals $\chi^2$ value by increasing $\Lm$. In general, the preferred fit for any given ansatz corresponds to the smallest $\Lm$ for which the goodness of the fit is reasonable and for which subsequent increases in $\Lm$ do not cause the $\chi^2$ value to drop by vastly more than one unit per degree of freedom. In practice, by “reasonable” we mean that $\chi^2/\rm{DF} \approx 1$, where DF is the number of degrees of freedom. The systematic error is estimated by comparing estimates from various sensible fitting ansatz.

\subsection{7D percolation with PBC}
\subsubsection{The fractal dimension} 
We start with discussing sizes of percolation clusters. As predicted from the CG asymptotics, we expect the size of the largest cluster $C_1 \sim L^{D_L}$ with the finite-size fractal dimension $D_L = 2d/3$. Indeed, in Fig.~\ref{fig:C1_ckra}(a), we log-log plot the data of $C_1$ versus $L$, and clearly the slope is consistent with $14/3$. To precisely estimate $D_L$, we perform the least squares fitting of the $C_1$ data to the ansatz~\eqref{eq:fitting_ansatz}. We first fit the data by including only $a_0$ and $a_1$ terms in the ansatz, i.e., fixing $a_2, c_0$ to 0. Even when $L_{\rm min} = 4$, the fitting becomes stable which estimates $y_{C_1} = 4.666(2)$ in perfect agreement with $14/3$ and the leading correction-to-scaling exponent $y_1 \approx -3$, implying that the $C_1$ data suffers quite weak finite-size corrections. Including a constant term $c_0$ to the fitting ansatz produces similar estimate of $y_{C_1}$, and the estimate of $c_0$ is consistent with zero. The results are summarized in Table~\ref{tab:fit_C1}. By comparing estimates from various reasonable ansatz, we estimate $y_{C_1} = 4.666(5)$, consistent perfectly with $D_L = 14/3$ at 7D.

We next study the thermodynamic fractal dimension $D_F$ of percolation clusters, from the scaling of cluster sizes $s$ versus the unwrapped radius of gyration $R$. In Fig.~\ref{fig:C1_ckra}(b), we plot in log-log scale the data of $s$ versus $R$, and our data clearly suggest that $s \sim R^{4}$, implying $D_F = 4$ for all percolation clusters (with sizes $\gg 1$) including the largest one.

\begin{figure}[t]
        \centering
        \includegraphics[scale=0.68]{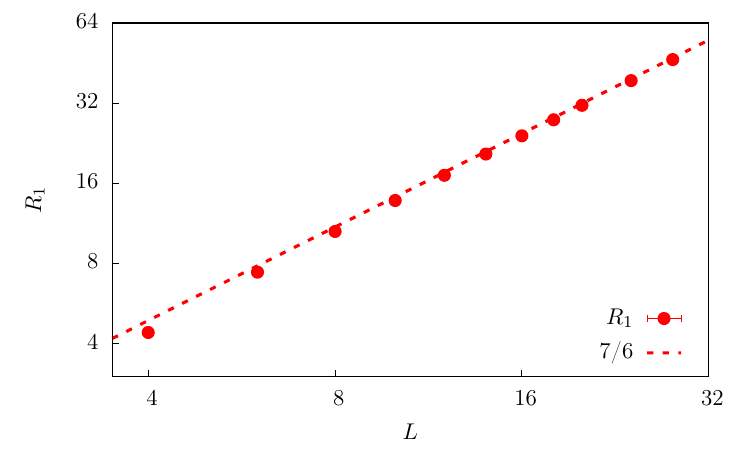}
        \caption{Plot of the unwrapped radius of gyration for the largest cluster $R_1$ versus $L$, which implies that $R_1 \sim L^{7/6}$.}
        \label{fig:R1}
    \end{figure}

In particular, we are interested in the scaling of $R_1$, which is the radius of gyration of the largest cluster and  effectively the unwrapped correlation length $\xi_u$ of the PBC case. By combining $C_1 \sim R^{D_F}_1$ and $C_1 \sim L^{D_L}$, one has $R_1 \sim L^{d/6}$. In Fig.~\ref{fig:R1}, we plot the data of $R_1$ versus $L$ in the log-log scale, and clearly the slope of data points is consistent with $7/6$. By a similar fitting procedure, we estimate $y_{R1} = 1.16(1)$, also consistent with $7/6$.


\begin{table}[b]
\centering
\begin{tabular}{|l|l|lllll|}
\hline 
$\scrO$ & $L_{\rm min}$  	&\quad \,$y_\scrO$\quad 	&\quad $a_0$ \quad	&\; $a_1$   &\; $y_1$\quad 	& $\chi^2/{\rm DF}$ 	\\
\hline 
& 4     &4.666(2)  	&1.140(7)  	&-6(1)       	&-3.1(2)   	&2.2/7\\ 
$C_1$ &6     &4.667(4)  	&1.14(1)   	&-10(16)    	&-3(1)     	&2.0/6\\ 
\cline{2-7}
& 6     &4.6658(9)  	&1.141(3)  	&-5.2(2)        &-3      	&2.2/7\\ 
& 8     &4.667(1)  	&1.137(5)  	&-4.5(7)        &-3  	   	&1.9/6\\ 
\hline 
\end{tabular} 
\caption{The fitting results for the size of the largest cluster of the critical 7D percolation with periodic boundary conditions.}
\label{tab:fit_C1} 
\end{table}



     \begin{figure}[t]
        \centering
        \includegraphics[scale=0.68]{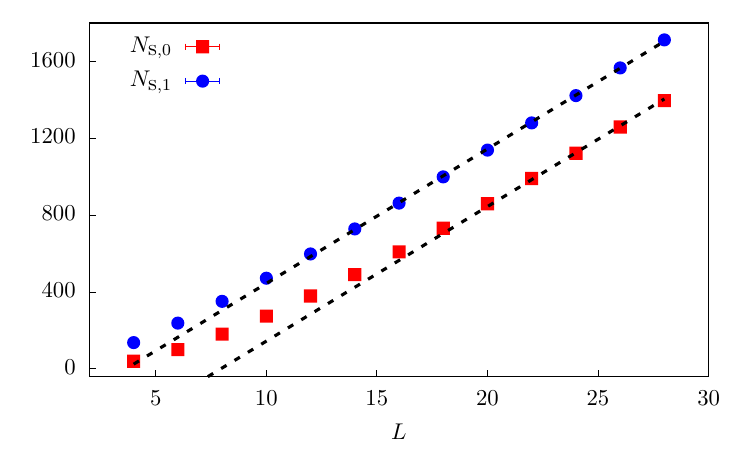}
        \caption{Plot of the number of spanning clusters $N_{\rm s,0}$ and $N_{\rm s,1}$ versus $L$. The dashed lines are straight lines, and thus it implies that $N_{\rm s,0}, N_{\rm s,1} \sim L$.}
        \label{fig:nsnc}
    \end{figure}  

\subsubsection{The number of spanning clusters}
In \cite{bennaim2005kinetic}, it was predicted that the cluster-number density $n(s,V)$ for the critical percolation on the CG follows  
\begin{equation}
n(s,V) \sim  s^{-5/2} \tilde{n}(s/V^{2/3})\;,
\end{equation}
which was numerically confirmed in Ref.~\cite{huang2018critical} on both the CG and 7D lattices with PBC. A cluster is called spanning if its unwrapped radius of gyration is at least $L$. Since the scaling $s \sim R^4$ holds for all percolation clusters, it immediately follows that a spanning cluster has size at least of order $L^4$. Then the number of spanning clusters can be calculated as
\begin{align}
\label{eq:nsd}
N_s &= L^d \int_{c L^4} s^{-5/2} \tilde{n}(s/L^{2d/3}) ds \sim  L^{d-6}\;, 
\end{align}
where $c$ is some positive constant. In the second step, we assume the term $\tilde{n}(s/L^{2d/3})$ is a constant in the integral range.

\begin{figure}[t]
    \centering
    \includegraphics[scale=0.68]{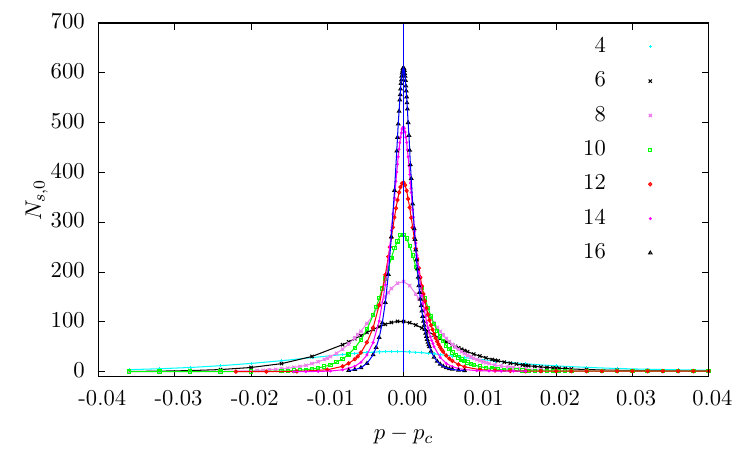}
    \caption{Plot of the number of spanning clusters $N_{\rm s, 0}$ near the critical point $p_c$. The values of $N_{\rm s, 0}$ on both sides of $p_c$ exhibit excellent symmetry.}
    \label{fig:nst}
\end{figure}

In simulation, we measure two number of spanning clusters $N_{\rm s,0}$ and $N_{\rm s,1}$, which are expected to have the same scaling behaviour but possibly with different finite-size corrections. In Fig.~\ref{fig:nsnc}, we plot the data versus $L$, which strongly suggests $N_{s,0}, N_{s,1} \sim L$. We then perform the least-square fitting to the ansatz~\eqref{eq:fitting_ansatz}. For both $N_{\rm s,0}$ and $N_{\rm s,1}$, we find that leaving $y_1$ free in the fitting ansatz cannot produce stable results. We then try to fit by fixing $y_1 = -1$ and $y_2 = -2$, and the fits estimate the leading exponent is $1.00(5)$ for $N_{s,0}$ and $1.00(3)$ for $N_{s,1}$, both consistent with the expected value $1$. We also study the number of spanning clusters near the critical point. As Fig.~\ref{fig:nst} shows, $N_{s,0}$ has a peak value at $p_c$ and it is surprisingly symmetric near $p_c$.
 

\subsubsection{\texorpdfstring{The susceptibility $\chi$ and $\chi_{\bf{k}}$}{FSS of the susceptibility chi and chik}}
\begin{figure}[t]
            \centering
            \includegraphics[scale=0.68]{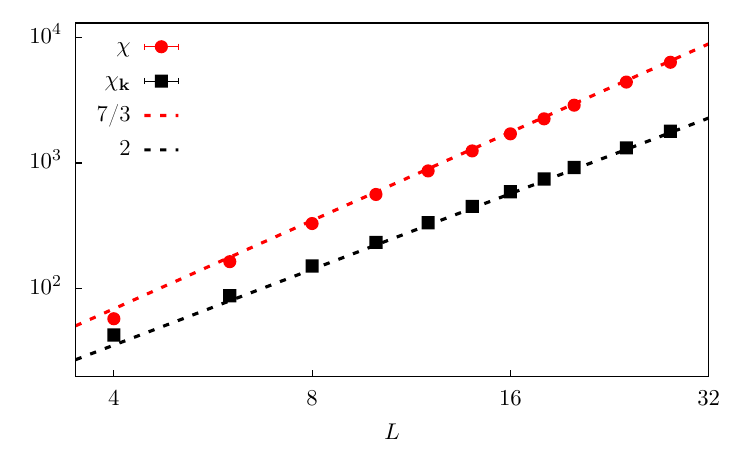}
            \caption{Log-log plot of the susceptibility $\chi$ and its Fourier mode $\chi_{\bf{k}}$ with non-zero $\bfk$. It implies that $\chi \sim L^{d/3}$ following the CG asymptotics, and $\chi_{\bfk} \sim L^2$ following the GFP prediction.}
            \label{fig:chi}
        \end{figure}

We next discuss the susceptibility $\chi$ and its Fourier transformation $\chi_{\bf{k}}$ with non-zero mode $\bf{k}$. In Fig.~\ref{fig:chi}, we plot the data of $\chi$ and $\chi_{\bf{k}}$ at the critical point in the log-log scale, and it suggests the scaling $\chi \sim L^{d/3}$ and $\chi_{\bf{k}} \sim L^2$; namely $\chi$ follows the CG asymptotics and $\chi_{\bf{k}}$ follows the GFP prediction. To estimate the leading scaling exponents, we fit the data of $\chi$ and $\chi_{\bf{k}}$ to the ansatz~\eqref{eq:fitting_ansatz}. For $\chi$, we first fit by fixing $a_2, c_0 = 0$ and the fitting is stable when $L_{\rm min} = 4$ and estimates $y_{\chi} = 2.337(4)$ and $y_1 = -2.5(1)$. We then try to include $c_0$ into the fitting ansatz,  but no stable results can be obtained. We also try the fitting by fixing $y_1$ to $-1$ or $-2$, and then similar estimates of $y_\chi$ are obtained. By comparing the results from various ansatz, we estimate $y_\chi  = 2.34(1)$, consistent with $7/3$. Similar analysis has been done for $\chi_{\bf{k}}$ and we finally estimate $y_{\chi_k} = 2.00(2)$, which is consistent with the expected value $2$.

\subsection{7D percolation with FBC}
\subsubsection{\texorpdfstring{The pseudo-critical point $p_L$}{The pseudo-critical point pL}}
\begin{figure}[t]
            \centering
            \includegraphics[scale=0.67]{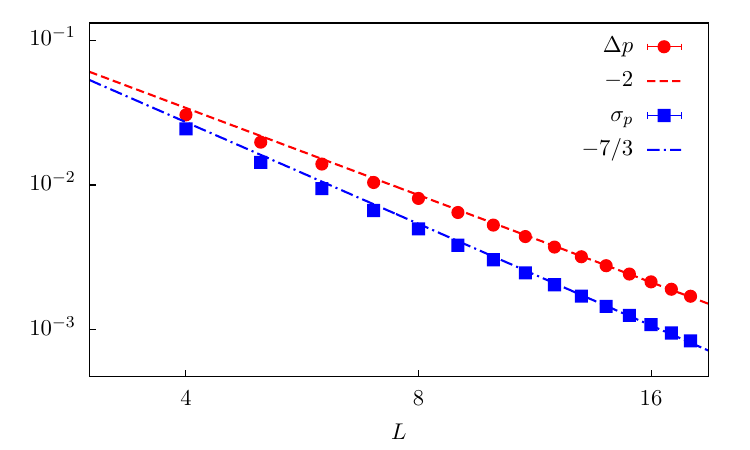}
            \caption{Log-log plot of $\Delta_p$ which is defined as $p_L - p_c$, and $\sigma_{p}$ which is the standard deviation of $\tilde{p}_L$. It implies that $\Delta_p \sim L^{-2}$ and $\sigma_{p} \sim L^{-d/3}$.}
            \label{fig:pL_he}
        \end{figure}

For the FBC case, we first study the pseudo-critical point $p_L$. In Fig.~\ref{fig:pL_he}, we plot $\Delta_p$ which is the distance between $p_L$ and the critical point $p_c$, and it suggests that $p_L - p_c \sim L^{-2}$ and $p_L > p_c$ is in the low-temperature region. We then perform the least-square fitting of the $p_L$ data to the ansatz~\eqref{eq:fitting_ansatz}. Leaving $c_0$ free cannot produce stable fits, so instead we fix $c_0$ at the known $p_c$ value. Including only the $a_0$ and $c_0$ terms to the ansatz leads to large residuals ($\chi^2$ value) of the fit, so we try to include the correction-to-scaling terms to the ansatz. However, leaving $y_1$ free still cannot produce stable fits, so we fix $y_1 = -1$ and $y_2 = -2$. Stable fits are obtained which estimate $y_{p_L} = -2.01(3)$ consistent with $-2$. Similar procedure is done for the standard deviation $\sigma_p$ and we finally estimate $y_{\sigma_p} = -2.3(2)$, which is consistent with $-d/3$ at 7D. Thus, for the FBC case, our data suggest that the pseudo-critical point is in the low-temperature phase, with distance of order $L^{-2}$ from $p_c$, and the critical window around $p_L$ is of order $L^{-d/3}$. Namely, $p_c$ is out of the critical region of $p_L$.


\subsubsection{The fractal dimension} 
   
\begin{figure}[t]
            \centering
            \includegraphics[scale=0.68]{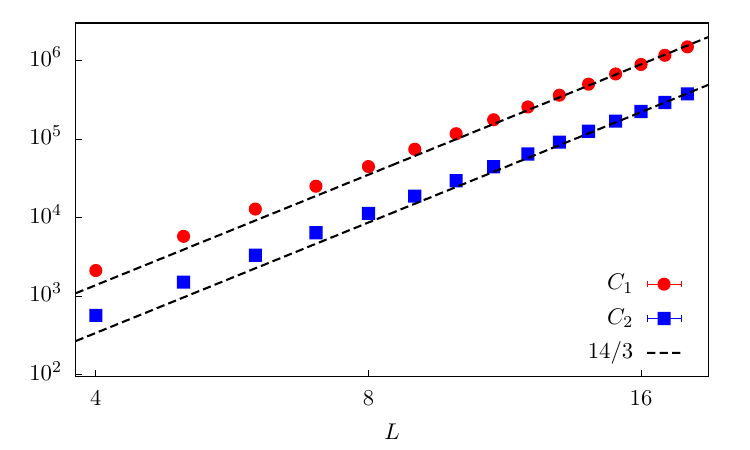}
            \caption{Log-log plot of the sizes of the largest cluster size $C_1$ and the second largest cluster $C_2$ at the pseudo-critical point $p_L$ of the FBC case. It implies that $C_1, C_2 \sim L^{2d/3}$, which is the same scaling as the PBC case at $p_c$.}
            \label{fig:C1}
        \end{figure}

We then measure the sizes of the two largest cluster size $C_1$ and $C_2$ at the pseudo-critical point $p_L$. In Fig.~\ref{fig:C1}, we plot in the log-log scale $C_1$ and $C_2$ versus $L$, and clearly our data suggest that both of them scale as $L^{2d/3}$. We then perform least-square fitting of the data of $C_1$ and $C_2$ to the ansatz ~\eqref{eq:fitting_ansatz}. Due to the strong finite-size corrections, we cannot obtain stable fits for $C_2$. But from $C_1$, the fitting results estimate $y_{C1} = 4.5(1)$, consistent with $14/3$ within two standard deviations. Thus, our data show that, for FBC at the pseudo-critical point $p_L$, the sizes of two largest clusters exhibit the same scaling as the PBC case at the critical point $p_c$.

\subsection{7D percolation with CBC}
\subsubsection{Correlation length}

\begin{figure}[t] 
\centering 
\includegraphics[scale=0.67]{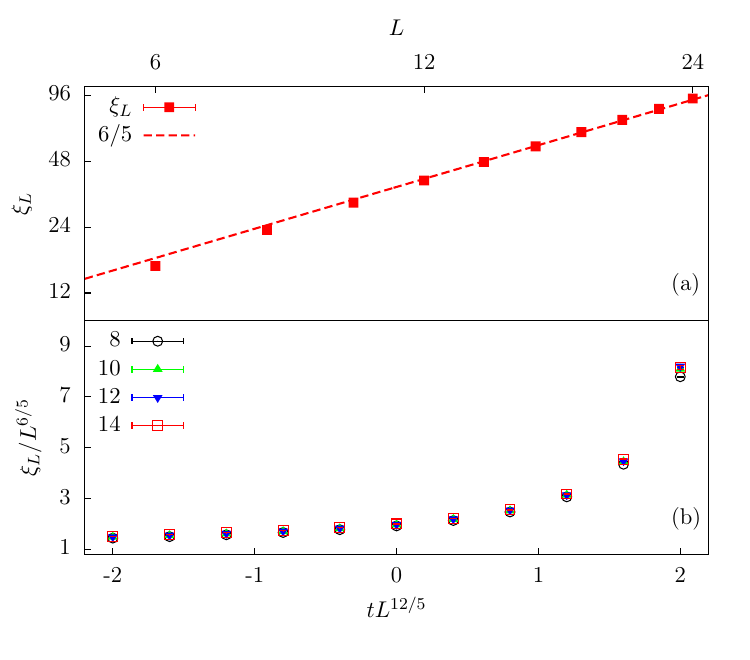} 
\caption{Plot of the correlation length $\xi_L$ of the CBC case. (a) Log-log plot of $\xi_L$ versus $L$ to show the scaling $\xi_L \sim L^{(d-1)/5}$ at the critical point. (b) Plot of $\xi_L$ in the critical window to show the universal scaling function. They strongly support the validity of Eq.~\eqref{eq:corre}} 
\label{Fig.corr_window} 
\end{figure}

We first study the correlation length $\xi_L$ along the axial direction, which is $\tau_{\rm max}$, the largest layer the cluster can grow up to. In Fig.~\ref{Fig.corr_window}(a), we plot the data of $\xi_L$ in the log-log scale, and the data points collapse onto a line with slope $6/5$ which suggests the scaling $\xi_L \sim L^{6/5}$. We then fit the data of $\xi_L$ to the ansatz~\eqref{eq:fitting_ansatz}. Including only one correction-to-scaling term to the ansatz and leaving $y_1$ free, we obtain stable fitting with $y_{\xi_L} = 1.19(1)$ and $y_1$ close to $-1$. Fixing $y_1 = -1$ produces similar results. By considering the systematic errors, we estimate $y_{\xi_L} = 1.19(2)$, consistent with the expected value $6/5$.



We then study the the FSS behavior of $\xi_L$ near the critical point. For each system size $L$, we simulate at several $p$ values such that $t L^{12/5}$ are constants in $[-2, 2]$ with $t=(p - p_c)/p_c$. We then plot $\xi_L L^{-6/5}$ versus $t L^{12/5}$, as shown in Fig.~\ref{Fig.corr_window}(b). The excellent data collapse provides strong evidence for the validity of Eq.~\eqref{eq:corre}.



\subsubsection{Correlation function}

Before discussing the correlation function in the cylinder, we first review some known results for the surface critical behaviour of percolation on the upper-half plane of $\ZZ^d$. Each edge, including these on the surface, is occupied with probability $p_c$, the critical point of the bond percolation model on $\ZZ^d$. The two-point correlation function perpendicular to the surface scales as $g_{\perp}(r) \sim r^{2-d-\eta_{\perp}}$, and the susceptibility, which is the mean size of finite clusters connecting to a site on the surface, scales as $\chi_{\perp} \sim \left|p - p_c \right|^{-\gamma_{\perp}}$. The surface order parameter $P$, defined as the probability that a chosen site on the surface belongs to the infinite cluster, scales as $P \sim (p - p_c)^{\beta_s}$. The standard scaling relation $\gamma_{\perp} = (2-\eta_{\perp})\nu$~\cite{de1980percolation} still holds, with $\nu = 1/2$ is the correlation-length critical exponent. The mean-field theory predicts that $\gamma_{\perp} = 1/2$~\cite{de1981mean} and $\beta_s = 3/2$. It implies that $\eta_{\perp} = 1$, which means $g_{\perp}(r) \sim r^{1-d}$.

\begin{figure}[t] 
\centering 
\includegraphics[scale=0.68]{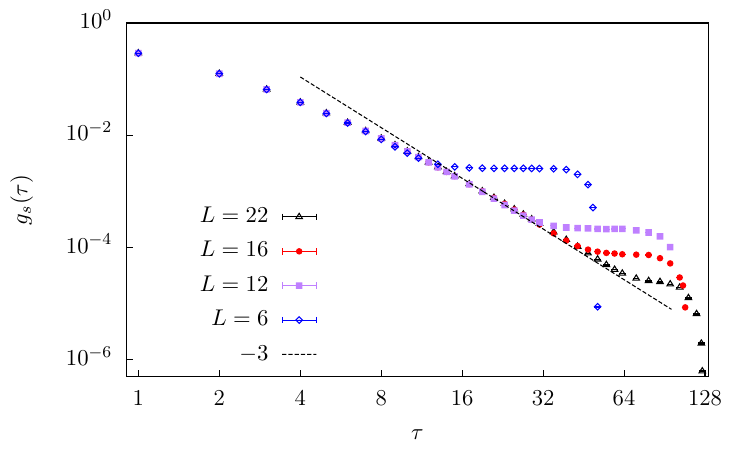} 
\caption{Plot of the one-point surface correlation function $g_s(\tau)$ versus $\tau$ for the CBC case. It implies that $g_s(\tau)$ first exhibits the power-law decay $\tau^{(1-d)/2}$ and then enters a plateau, and finally decays significantly fast.} 
\label{Fig.gr} 
\end{figure}

\begin{figure}[t] 
\centering 
\includegraphics[scale=0.67]{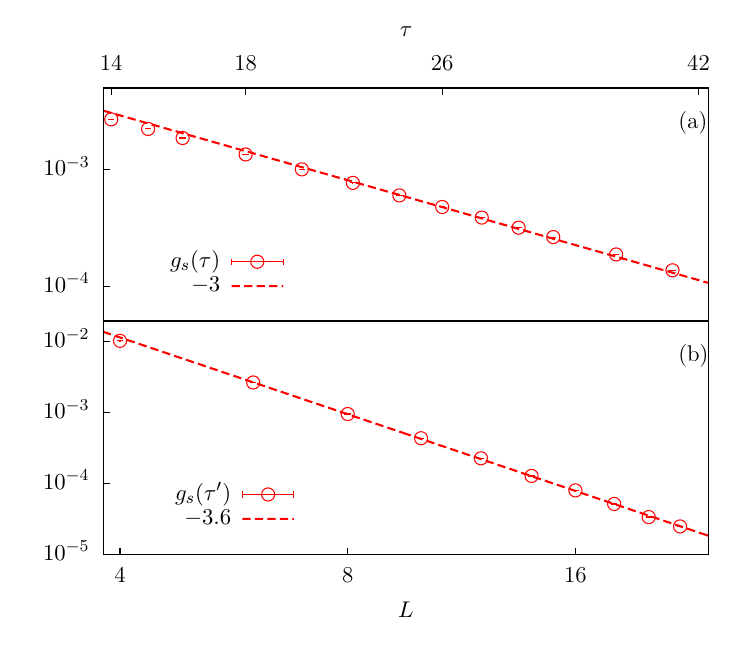} 
\caption{Plot of the one-point surface correlation function $g_s(\tau)$ in the power-law region (top) and in the plateau region (bottom), for the CBC case. It implies that $g_s(\tau) \sim \tau^{(1-d)/2}$ for short distance and then enters the plateau $g_s(\tau') \sim L^{-18/5}$ when $\tau' \sim \xi_L$ at 7D.} 
\label{Fig.gr_double} 
\end{figure}

We next study the one-point surface correlation function $g_s(\tau)$, which is defined as the probability that a site at layer $\tau$ belongs to the cluster growing from the fully-occupied initial layer. Since $g_s(\tau)$ is the one-point correlation function, we expect $g_s(\tau) \sim \tau^{(1-d)/2}$. We next plot the data of $g_s(\tau)$ versus $\tau$ in the log-log scale in Fig.~\ref{Fig.gr}, and clearly there are three different regions. One can see $g_s(\tau)$ first exhibits the power-law decay $\tau^{(1-d)/2}$, and then enters a plateau before it decays significantly fast. If $g_s(\tau)$ enters the plateau when $\tau \sim \xi_L \sim L^{(d-1)/5}$, then $g_s(\xi_L) \sim L^{-(d-1)^2/10}$ which is $L^{-18/5}$ at 7D. To numerically confirm this, we choose $\tau^\prime = \frac{6}{5} L^{(d-1)/5}$ such that for the system sizes we simulated, $\tau^\prime$ is in the plateau region, and study the scaling of $g_s(\tau^\prime)$. In Fig.~\ref{Fig.gr_double}, we plot $g_s(\tau^\prime)$ versus $L$ in the log-log scale, and clearly the data points collapse onto a straight line with slope $-18/5$.

Using the scaling of $\xi_L$ in Eq.~\eqref{eq:corre}, one can derive the FSS formula for other quantities. For example, let $P(t, L)$ be surface order parameter of a finite box. It follows by Eq.~\eqref{eq:corre}, Eq.~\eqref{Eq:PBCtoGFP} and the scaling $P \sim (p - p_c)^{\beta_s}$ that $P\left(t, L\right)=L^{-3(d-1)/5} \tilde{P} \left[tL^{2(d-1)/5}\right]$. The scaling at the critical point $t=0$ can be used to explain the emergence of plateau in $g_s(\tau)$. When $\tau$ is small, the layer $\tau$ is strongly influenced by the initial fully-occupied layer and thus $g_s(\tau)$ is larger than $P(0,L)$. As $\tau$ increases, $g_s(\tau)$ decays as $\tau^{(1-d)/2}$. But, as long as the largest cluster at layer $\tau$ is the same cluster growing from the initial layer, then $g_s(\tau)$ is at least $P(0,L)$, which corresponds to the plateau. When $\tau$ is large enough such that the largest cluster at layer $\tau$ is not the cluster growing from the initial layer, then $g_s(t)$ decays significantly fast.

We then fit the data of $g_s(\tau)$ in the power-law region. The power-law decay of $g_s(\tau)$ is valid only for large $\tau$, but when $\tau$ is too large $g(\tau)$ is affected by the plateau. We thus choose the data of $L=22$ to fit and set the fitting range as $\tau \in \left[16, 40\right]$. We cannot obtain stable fitting results when let $y_1$ free, and by fixing $y_1 = -1$ the fit is stable which estimates the slope to be $3.00(2)$, consistent with the predicted value $(1-d)/2$ at 7D. We show the power-law region of $g_s(\tau)$, using the data from the system size $L=22$, in Fig.~\ref{Fig.gr_double}.  


\section{Discussion}
\label{Sec:Discussion}
In this paper, we systematically studied the bond percolation model on seven-dimensional hypercubic lattices with periodic (PBC), free (FBC) and cylindrical boundary conditions (CBC). Our results show that, above the upper critical dimension, the finite-size scaling of percolation is determined by the interplay of the Gaussian fixed-point (GFP) predictions and the complete-graph (CG) asymptotics. For PBC, the GFP determines the distance-dependent properties, such as the short-distance behaviour of the two-point function and the Fourier transformed susceptibility with non-zero modes. The CG asymptotics dominates the scaling behaviour of macroscopic quantities with respect to the linear system size $L$, such as the size the largest cluster and susceptibility. The existence of the CG asymptotics is due to that, with periodic boundaries, the characteristic length of large percolation clusters, which is the unwrapped radius of gyration, diverges faster than $L$. If one uses the unwrapped radius as the length scale of percolation clusters, then the effect of periodic boundaries will be correctly taken into account and consequently the CG asymptotics are effectively hidden. At the percolation threshold, there exist three magnetic-type exponents. The thermodynamic fractal dimension $D_F=4$ describes the dependence of the size of a cluster on its unwrapped gyration radius as $C \sim R^{D_F}$; the finite-size fractal dimension $D_L=2d/3$ is for the $L$-depdence of the largest-cluster size as $C_1 \sim L^{D_L}$; the RG magnetic exponent $y_h=1+d/2$ is for $\chi_{\bfk} \sim L^{2y_h-d}$ for non-zero momentum $\bfk$. Note that, for low spatial dimensions $d<6$, the three exponents are identical and there are only one magnetic exponent. For FBC, since the correlation length is cut off by $L$, at the critical point the finite-size scaling follows the GFP predictions. However, we provided strong numerical evidence that in the low-temperature scaling window, some quantities exhibit the CG asymptotics, such as the sizes of the largest and second largest clusters. 

More interesting interplay can be seen in the CBC case, where one dimension is infinite and the other $d-1$ dimensions are periodic. Along the infinite dimension, the interplay of the effects from GFP and CG leads to a correlation length $\xi_L \sim L^{(d-1)/5}$, different scaling with both the PBC and FBC cases. We provided a field-theoretical calculation and also numerical evidence to support the scaling of $\xi_L$. We also measured the one-point surface correlation function $g_s(\tau)$. The numerical data show it first exhibits the mean-field prediction $\tau^{(1-d)/2}$ and then enters a plateau of order $L^{-18/5}$ when $\tau \sim \xi_L$. It decays exponentially fast when $\tau\gg \xi_L$.

Some interesting but open questions are worth discussing here. In the FBC case, the behaviour of the two-point function in the low-temperature scaling window is still worth investigating. Although the length scale (standard or unwrapped) cannot exceed $L$ with FBC, it is still possible for the two-point function to exhibit the piece-wise behaviour, if the percolation clusters are large enough. It is also interesting to extend the CBC study to a more general case. Namely, for a $d$-dimensional lattices with $d\geq d_c$, let the boundaries in $d^\prime$ dimensions be open and the other $d-d^\prime$ dimensions periodic. One can systematically study the finite-size scaling as $d^\prime$ increases from 1 to $d-1$. In any cases, we believe that their finite-size scaling behaviors can be understood from the perspective of the interplay between the CG and GFP asymptotics. In particular, by properly defining an effective volume $V_{\rm eff}$, there exists a (unwrapped) length scale $\xi_u  \sim V_{\rm eff}^{1/6}$. In terms of $\xi_u$, the largest-cluster size scales as $\xi_u^4$ and the finite-size critical window is $\delta p\sim \xi_u^{-2}$. 
  





\section*{Acknowledgements}
This work has been supported by the National Natural Science Foundation of China (under Grant No.12275263), the National Key R\&D Program of China(under Grant No. 2018YFA0306501).

 \bibliographystyle{apsrev4-2}

\end{document}